\documentclass[aps,pra,reprint,superscriptaddress]{revtex4-2}

\usepackage[english]{babel}
\usepackage{graphics}
\usepackage{graphicx}
\usepackage{amsfonts,amssymb,amsmath}
\usepackage{color}
\usepackage[normalem]{ulem}

\definecolor{darkblue}{rgb}{0.0,0.0,0.4}
\definecolor{darkgreen}{rgb}{0.0,0.4,0.0}
\definecolor{darkred}{rgb}{0.6,0.0,0.0}
\usepackage[bookmarksopen]{hyperref}
\hypersetup{colorlinks,linkcolor=darkblue,citecolor=darkblue,urlcolor=darkblue}

\begin{document}

\title{Microscopic 3D printed optical tweezers for atomic quantum technology}
\author{Pavel Ruchka} 
\thanks{These two authors contributed equally}
\affiliation{4. Physikalisches  Institut, Research Center SCoPE and  Center  for  Integrated  Quantum  Science  and  Technology, Universit\"at  Stuttgart,  Pfaffenwaldring  57,  70569  Stuttgart,  Germany}

\author{Sina Hammer}
\thanks{These two authors contributed equally}
\affiliation{5. Physikalisches  Institut  and  Center  for  Integrated  Quantum  Science  and  Technology,Universit\"at  Stuttgart,  Pfaffenwaldring  57,  70569  Stuttgart,  Germany}

\author{Marian Rockenh\"auser}
\affiliation{5. Physikalisches  Institut  and  Center  for  Integrated  Quantum  Science  and  Technology,Universit\"at  Stuttgart,  Pfaffenwaldring  57,  70569  Stuttgart,  Germany}

\author{Ralf Albrecht}
\affiliation{5. Physikalisches  Institut  and  Center  for  Integrated  Quantum  Science  and  Technology,Universit\"at  Stuttgart,  Pfaffenwaldring  57,  70569  Stuttgart,  Germany}

\author{Johannes Drozella}
\affiliation{Institute of Applied Optics (ITO) and Research Center SCoPE, University of Stuttgart, Pfaffenwaldring 9, 70569 Stuttgart, Germany}

\author{Simon Thiele}
\affiliation{Institute of Applied Optics (ITO) and Research Center SCoPE, University of Stuttgart, Pfaffenwaldring 9, 70569 Stuttgart, Germany}

\author{Harald Giessen}
\affiliation{4. Physikalisches  Institut, Research Center SCoPE and  Center  for  Integrated  Quantum  Science  and  Technology, Universit\"at  Stuttgart,  Pfaffenwaldring  57,  70569  Stuttgart,  Germany}

\author{Tim Langen}
\email{t.langen@physik.uni-stuttgart.de}

\affiliation{5. Physikalisches  Institut  and  Center  for  Integrated  Quantum  Science  and  Technology,Universit\"at  Stuttgart,  Pfaffenwaldring  57,  70569  Stuttgart,  Germany}

\begin{abstract}
Trapping of single ultracold atoms is an important tool for applications ranging from quantum computation and communication to sensing. However, most experimental setups, while very precise and versatile, can only be operated in specialized laboratory environments due to their large size, complexity and high cost. Here, we introduce a new trapping concept for ultracold atoms in optical tweezers based on micrometer-scale lenses that are 3D printed onto the tip of standard optical fibers. The unique properties of these lenses make them suitable for both trapping individual atoms and capturing their fluorescence with high efficiency. In an exploratory experiment, we have established the vacuum compatibility and robustness of the structures, and successfully formed a magneto-optical trap for ultracold atoms in their immediate vicinity. This makes them promising components for portable atomic quantum devices. 
\end{abstract}

\maketitle


\section{Introduction}
Starting from the early landmark work on the trapping of dielectric particles~\cite{AshkinDielectric,AshkinRadiationPressure}, optical tweezer traps have matured into one of the key technologies for the manipulation of single, ultracold atoms in quantum science and technology~\cite{Kaufman2021}. Applications range from quantum simulation~\cite{Bernien2017,Leseluc2019,Scholl2021,Ebadi2021} and computing~\cite{Kaufman2015,Levine2019,Madjarov2020} to studies of fundamental chemical processes and collisions~\cite{Liu2018,Cheuk2020,Reynolds2020}, precision timekeeping~\cite{Madjarov2019,Young2020}, interfacing with nanophotonic structures~\cite{Thompson2013,Kim2019} and single-photon generation~\cite{Garcia2013}. 

However, experimental setups combining optical tweezers and ultracold atoms typically have macroscopic dimensions, are complex, and require sophisticated calibration and stable laboratory environments. In order to realize more compact, robust, and user-friendly technological applications, it is thus highly desirable to simplify and further miniaturize these setups. 

\begin{figure}[tb]
\vspace{8pt}
    \includegraphics[width=0.50\textwidth]{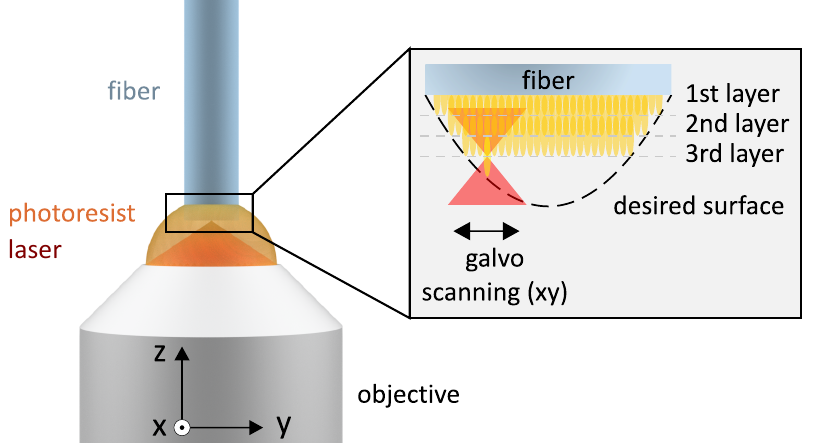}
    \caption{Schematic representation of the printing process using two-photon polymerization. We use a dip-in method, where a drop of photoresist is placed onto a microscope objective. The fiber is then dipped into this drop and a single voxel of the final structure is created in the focal spot of a high-power writing laser. Arbitrary structures are constructed layer by layer from many individual voxels, by scanning the laser focus position both laterally 
    (xy) using a galvo scanner and vertically (z) by displacing the objective.}
    \label{fig:fabrication}
\end{figure}

Here, we introduce a new optical tweezer concept for ultracold atoms that can meet these requirements. Our approach is based on microscopic lenses that are 3D printed on the tip of optical fibers. Similar 3D printed structures have already been used to realize a diverse range of optical components from endoscopes~\cite{Li2020} to multi-lens objectives~\cite{Gissibl2016}. This enabled the realization of tweezer traps for classical microscopic objects in chemical and biological environments~\cite{Asadollahbaik2020,Asadollahbaik2022}, large-size microlens arrays~\cite{Schaffner2020} and the coupling of single quantum dot emission into an optical fiber~\cite{Sartison2021}. In the following, we establish that these micrometer-sized structures are also well suited to manipulate ultracold atoms inside an ultrahigh vacuum environment. In particular, due to the large flexibility afforded by the 3D printing, it will be possible to integrate single-atom trapping \textit{and} high-fidelity detection into a single, compact device. 

\section{Fabrication}
3D printing has revolutionized many areas of research, engineering and production, making it one of the most significant inventions of recent years~\cite{Murr2016}. With the existing techniques it is possible to produce individually designed objects from a large variety of materials, with sizes ranging from micro- to macro-scales~\cite{Do2015,Yi2017,Saint2018,Chen2019,materials_Schmid:19,macro_Ristok:20}.

3D printing of micrometer-sized optical components is possible using two-photon polymerization~\cite{Serbin2003,Deubel2004, 2PP_general, 2PP_approaches}. In this technique, the desired structures are directly written into a photoresist using a focused laser beam (Fig.~\ref{fig:fabrication}). For the printing of microscopic fiber lenses, we use negative photoresists, which polymerize after exposure~\cite{Quero2018}. The focal spot of the laser beam defines the smallest building block of any solid structure formed in this way. This smallest unit is commonly referred to as a \textit{voxel}. 

We use a commercially available 3D printer~\cite{nanoscribe}, together with a pulsed femtosecond laser at $780\,$nm and pulse powers of around $2.25\,$nJ to realize voxels with dimensions down to $100\,$nm. This minimum voxel size sets the scale for the design of any larger structures~\cite{Li2020}.

The lenses are then built sequentially out of such voxels using a dip-in laser lithography configuration. In this configuration, the photoresist is directly placed onto the objective for the writing laser. For the writing process, a cleaved and oxygen plasma-activated fiber is dipped into the resist and the laser focus position is scanned using two galvo-mirrors and piezo actuators that move the objective. We increase the adhesion of the printed structure to the fiber by silanizing the substrate \cite{silanization}. To precisely align the fiber, the writing laser is attenuated and its position on the fiber is tracked using a CCD camera. After writing, the micro-optics are developed by placing them in a bath with developer~\cite{developper} for $15$~minutes and by rinsing them with a solvent. This dissolves the non-exposed parts of the negative photoresist to reveal the final,  printed structure. Finally, the micro-optics are UV-cured, which ensures full polymerization and, as a consequence, consistency of the refractive index.

The specific geometries of the lenses are designed using optical ray tracing~\cite{raytracing} and wave propagation method (WPM) simulations~\cite{Brenner:93, Schmidt:17}. Subsequently, the designs are translated into trajectories for the writing process. In this process, the mechanical configuration of the printing setup facilitates precise control of the voxel position in three dimensions. This allows us to adjust parameters like the voxel layer thickness along the fiber axis, as well as the distance between two layers in one printing plane according to the optical design of the desired structure and photoresist type. 
The resolution of the overall process is given by an interplay of several parameters, which include the magnification and numerical aperture of the objective, photopolymer characteristics, laser power, scanning speed and the precision of the optomechanical parts, such as galvo mirrors and positioning stages. As we demonstrate in the following, taking all of these aspects into account, we are able to closely reproduce the initial optical designs.

\begin{figure*}[tb]
\centering
\includegraphics[width=\textwidth]{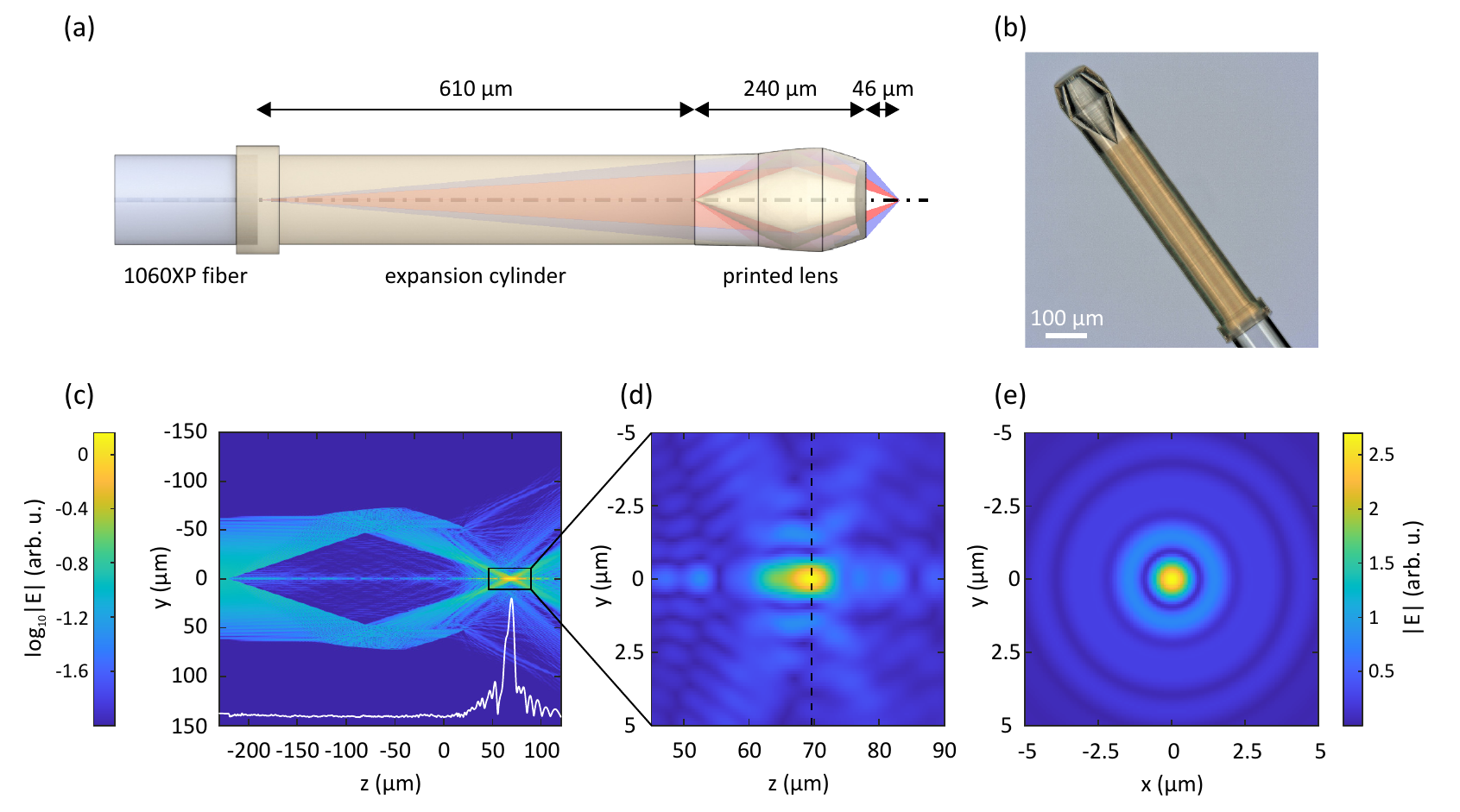}
\caption{TIR lens design. A sketch of the design is depicted in (a), highlighting the reflective, rather than diffractive nature of the design. (b) displays the printed lens structure under a microscope. In (c) the WPM-simulated electric field is shown on a logarithmic color scale. The white line is a cut along $z$ through the origin ($y=0$). The inset (d) corresponds to a zoom into the focal region of this simulation, revealing a high quality focal spot, with a $1/e$ diameter of $1.4\,\mu$m. (e) is the radial cross-section at the focal spot, as indicated by the dashed line in (d). The latter two panels are plotted using a linear color-scale. }
\label{fig:TIR_design}
\end{figure*}

\section{Example designs}
The goal of our concept is to form an optical tweezer by guiding light through a single mode optical fiber and focusing it down using the 3D printed optics. At the same time, resonant fluorescence emitted by atoms trapped in the tweezer is to be collected using the printed optics and, after separation from the incoming trapping light, guided towards a photodetector. 

Different approaches for the lens design are possible to realize this configuration. For the creation of a tweezer trap and, subsequently, efficient detection of the atomic fluorescence, a lens with a high numerical aperture (NA) is desirable. In the following, we present three examples for suitable lenses. Their key properties are summarized in Tab.~\ref{table:focal_shift}. Due to the flexibility afforded by the 3D printing process, many others --- tailored to meet specific requirements --- are conceivable.

\begin{table}[tb]
\centering
\begin{tabular}{l | c | c | c } 
     Lens type & Aspheric & Aspheric & TIR \\
     \hline\hline
     Numerical aperture & 0.45 & 0.6 & 0.75 \\
     Focal spot diameter & 2$\,\mu$m& 1.5$\,\mu$m & 1.4$\,\mu$m\\
     Working distance & $83.3\,\mu$m & $36.2\,\mu$m & $46.3\,\mu$m \\
     Chromatic focal shift & $-1.1\,\mu$m & $-0.4\,\mu$m & $-1.6\,\mu$m \\
     Solid angle covered & $4.83\%$ & $9\%$ & $15\%$
\end{tabular}
\caption{Parameters of the fiber lens examples presented in this work. The chromatic focal shift is given between the two wavelength of interest: 1064 and 780~nm. The working distance is given for 1064 nm and refers to the closest distance between the lens structures and the focal point (see Figs.~\ref{fig:TIR_design} and \ref{fig:asph_design} for reference). }
\label{table:focal_shift}
\end{table}

\subsection{Total-internal reflection design}
As a first example, we present a lens based on a total-internal reflection design (TIR), which is shown in Fig.~\ref{fig:TIR_design}. A similar design has previously been used to trap $\mu$m sized particles in water~\cite{Asadollahbaik2022}. 

To realize small foci suitable for single-atom trapping~\cite{Schlosser2001}, the light from the fiber is first expanded through a $610\,\mu$m long printed cylinder before subsequently passing through the lens. The lens itself uses fully reflective, rather than standard refractive optics, to achieve an NA of $0.75$ and a working distance of $46.3\,\mu$m. Larger NAs approaching unity are possible at the expense of a shorter working distance. One crucial benefit of such a lens for the work with ultracold atoms is the combination of a high NA with a large working distance, as can be seen by comparing it to more conventional designs with similar NA (see Table~\ref{table:focal_shift}). Additionally, the lens design could be optimized to achieve even lower chromatic focal shifts due to its reflective nature.

We simulate the resulting field using WPM simulations and find a high-quality focal spot with a $1/e$ diameter of 1.4$\,\mu$m. Printing is realized using commercially available Nanoscribe IP-Dip resist and takes approximately $5$ hours including the expansion cylinder. 

\begin{figure*}[tb]
\centering
\includegraphics[width=\textwidth]{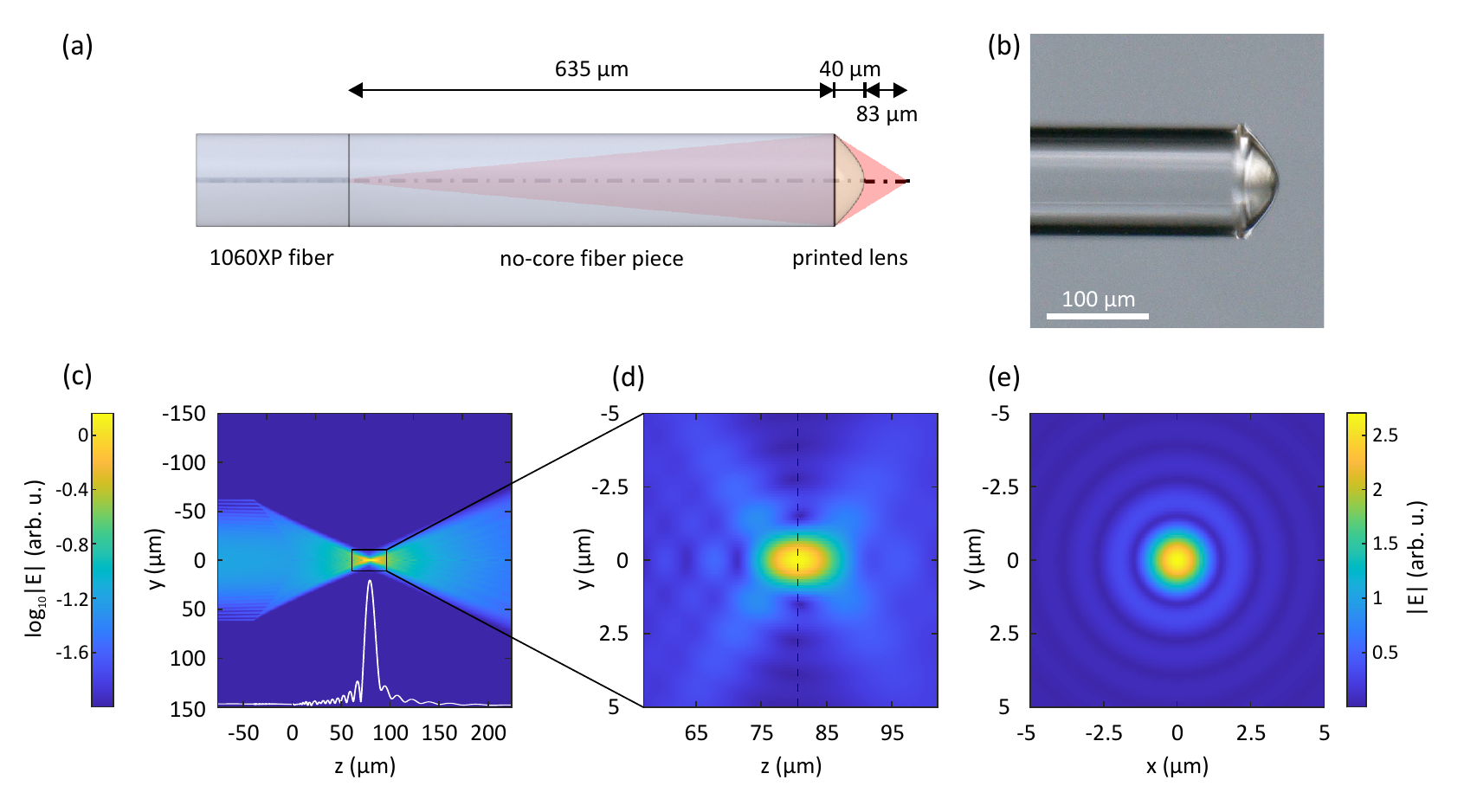}
    \caption{Aspherical lens design. (a) Sketch of the design for NA$=0.45$. (b) Microscope image of the corresponding lens. (c) WPM-simulations of the electric field, plotted using a logarithmic color-scale for clarity. The white line is a cut along $z$ through the origin ($y=0$). (d) Zoom into the focal region, with a $1/e$ diameter of $2\,\mu$m. (e) is the radial cross-section at the focal spot, as indicated by the dashed line in (d). The latter two panels are plotted using a linear color-scale.}
    \label{fig:asph_design}
\end{figure*}

\subsection{Aspheric design}
As a next example, we realize two types of aspheric lenses. Aspheric lenses are another possibility to achieve a high-quality focal spot using small-scale optics without the use of objectives containing multiple lenses. 

For our designs, we first expand the beam from a single-mode fiber in a $635\,\mu$m long expansion cylinder formed by a piece of no-core fiber that is spliced to the single-mode fiber. Subsequently, we focus down the beam using the respective 3D printed aspheric lens. Compared to the printed expansion cylinder this approach enables much faster production. The additional refraction at the cylinder-lens interface can be neglected for the design of the simple aspheric lenses. In principle, a similar approach could also be taken for the TIR optics discussed above, but due to angle mismatches that can be caused by the interface, this would require a more extensive optimization.

With the first lens design a NA of $0.45$ and a working distance of $83.3\,\mu$m are realized (see Fig.~\ref{fig:asph_design}). The second lens follows the same design principles and is characterized by an NA of $0.6$ and a working distance of $36.2\,\mu$m. We find focal spots with $1/e$ diameters of 2$\,\mu$m and 1.5$\,\mu$m, respectively. 

This highlights that the aspheric singlet design is an effective tool to create a small foci at a particular distance. However, while this solution is easy to design, simulate and implement, it has some limitations due to its refractive nature. For instance, the interdependence of NA, working distance and the size of the focal spot makes it challenging to create small foci at a higher distance, when one is limited by a $125\,\mu$m fiber diameter. However, it is possible to splice a no-core fiber with a larger diameter (e.g., $250\,\mu$m), which would allow for an increase of the working distance, while maintaining the relatively high NA and tight focusing.

As the structure of both aspheric designs is comparably simple, printing using Nanoscribe IP-S resist can be achieved in less than an hour for both of them. Compared to the IP-Dip resist used for the TIR lens, IP-S allows achieving smoother surfaces, which is of a special importance when printing the refractive aspheric optics.

\subsection{Detection efficiency}
To ensure that the fluorescence emitted by the atoms trapped in the focal spot can be collected and coupled back into the fiber, the focal spots for the different wavelengths used for trapping and fluorescence detection must be precisely matched. In the following, we will use rubidium atoms as an example, where fluorescence is emitted at $780\,$nm and a typical, far-detuned trapping wavelength is $1064\,$nm. All designs can easily be adjusted to cover also the wavelengths for a large variety of other atomic and molecular species.

We theoretically investigate the behavior of both lens designs by considering a trapped atom as a perfect source radiating uniformly in all directions and reversing the optical system using ray tracing software~\cite{raytracing}. Independent of the lens design we find that the focal spots for $1064\,$nm and $780\,$nm are displaced by $1-2\,\mu$m, and thus located exceptionally close to each other. The lenses thus cover between $5.4\%$ to $17\%$ of the solid angle of a trapped particle. As mentioned above, in principle, NAs of up to $0.99$ are possible, corresponding to detection efficiencies that can well exceed $40\%$.

Another important aspect for the detection efficiency of single trapped atoms are spurious photons that are created in the fiber or the polymer~\cite{Garcia2013}. Monitoring the output of the fibers with single-photon counters we have not observed any Raman shifted light at $780\,$nm from a $1064\,$nm laser beam for input powers up to $100\,$mW. We therefore expect the fluorescence detection based on the fiber lenses to be suitable down to the single-atom and single-photon level, even in the presence of significant powers for the realization of the tweezer trap. 

\subsection{Focal spot quality and trapping potentials}
We experimentally confirm our simulations, the optical properties of the fiber lenses and the quality of the 3D printing process using through-focus measurements. 

\begin{figure}[tb]
    \centering
    \includegraphics[width=0.45\textwidth]{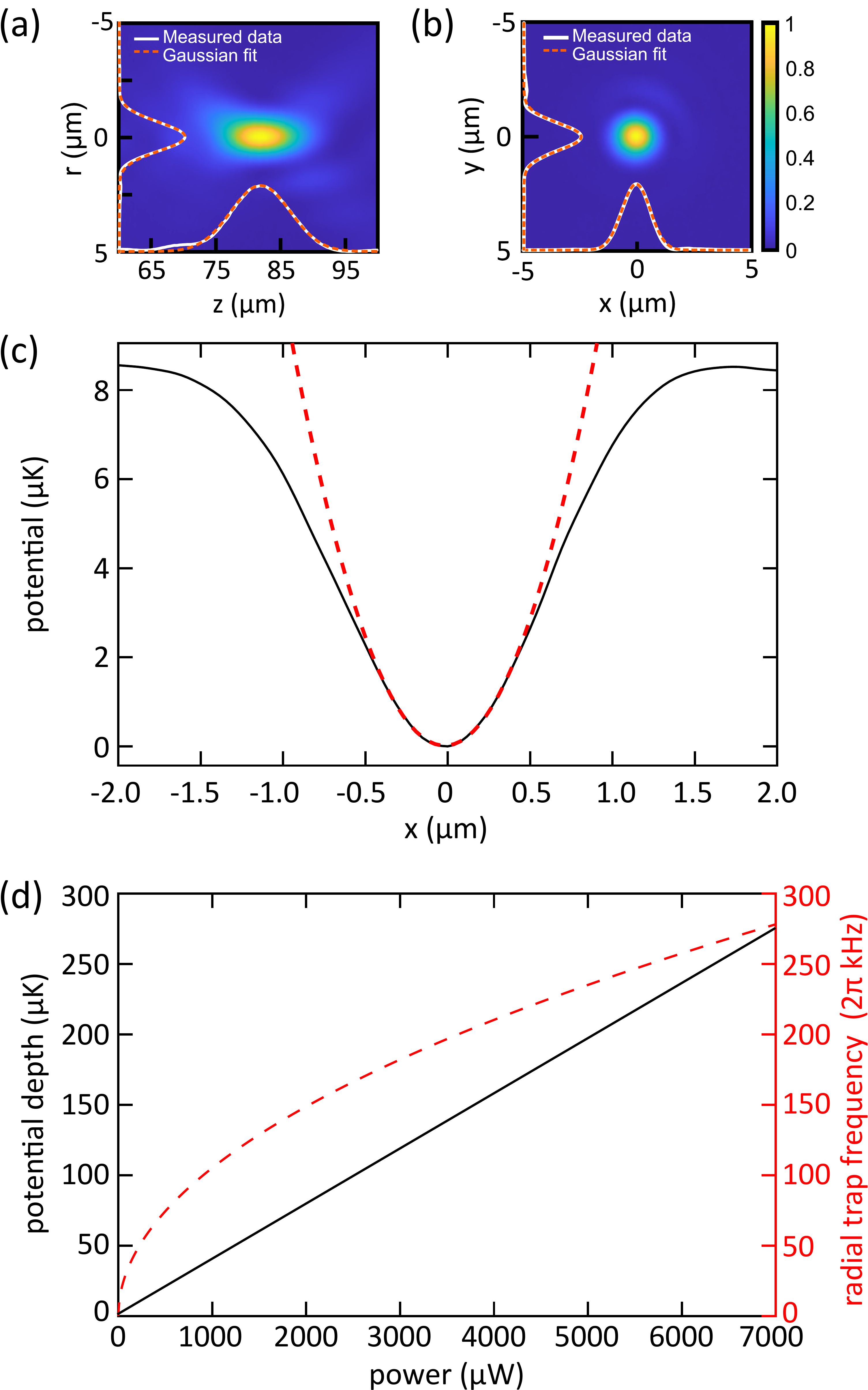}
    \caption{Measurements of the focal spot of the aspherical lens design with NA~=~0.45 along the laser beam axis (a) and perpendicular to it (b). Colors denote the normalized intensity. For this measurement, the power exiting the fiber was $220\,\mu$W. The result is in excellent agreement with the corresponding simulation shown in Fig.~\ref{fig:asph_design}d,e. (c) Examplary trapping potential along the perpendicular axis deduced from these measurements using Eq.~\ref{eq:potential}. The red dashed line is a harmonic fit, which yields a radial trapping frequency of approximately $\omega=2\pi\times\,73\,$kHz. The trapping potential shown is a cut along the horizontal dotted line. (d) Potential depth (solid black line) and radial trapping frequency (red dashed line) as a function of the power exiting the fiber. }
    \label{fig:potentials}
\end{figure}

An example, based on the aspherical lens design with NA$\,=0.45$, is shown in Fig.~\ref{fig:potentials}a,b. We find excellent agreement between our simulations and measurements, both in terms of working distance and focal spot, with a deviation of less than $2\%$ from the initial design.

In Fig.~\ref{fig:potentials}c,d, we further use this measurement to estimate the expected trapping potentials experienced by the atoms for a given input laser power. 

For alkali atoms, such as rubidium, the trapping potential $V(\mathbf{r})$ in a far-detuned tweezer trap can be approximated by a semi-classical model~\cite{Grimm2000}. This yields

\begin{equation}
    V(\mathbf{r})= \frac{3\pi c^2}{2 \omega_0^3}\frac{\Gamma}{\Delta} I(\mathbf{r}), \label{eq:potential}
\end{equation}
where $\omega_0$ is the atomic transition frequency, $\Gamma$ is the atomic linewidth, $\Delta$ is the detuning of the trapping light from the atomic transition, $c$ denotes the speed of light, and $I(\mathbf{r})$ is the position-dependent intensity distribution created by the fiber lens. For red detuning ($\Delta<0$) the dipole force $\mathbf{F} = \mathrm{-} \nabla V_\mathrm{dip}$ attracts atom towards the high intensity focal region. In our example system, rubidium, the remaining parameters are given by $\omega_0\sim 2\pi\times 384\,$THz, corresponding to a wavelength of $780\,$nm, $\Gamma=2\pi\times6\,$MHz, and $\Delta\sim2\pi\times 100\,$THz for trapping light at $1064\,$nm \cite{SteckRubidium}. Trapping potentials for more complex atoms with significant vector and tensor polarizabilities can be derived in a similar way, by combining vector WPM methods for the electric fields with precisely known data for the polarizabilities~\cite{Ravensbergen2018,Westergaard2011,LeKien2013}. 

Based on this, we find that already moderate input laser powers on the order of a few milliwatts generate traps that are hundreds of microkelvins deep, and thus sufficient to capture atoms from a typical magneto-optical trap. We analyze the trapping potentials further by extracting their characteristic frequencies from a harmonic approximation at the trap center. The results are on the order of hundreds of kHz. We conclude that the very small trapping volume and steep traps are well suited to realize sub-Poissonian loading of single atoms via light-induced collisions~\cite{Schlosser2001}. 

\section{Compatibility with ultracold atom technology}
To establish the compatibility of the printed structures with ultracold atom technology, we demonstrate the operation of a magneto-optical trap near the printed structures. This allows us to specifically test several crucial aspects of our concept: 
\setlength\leftmargini{10pt} 
\begin{itemize}
    \item First, the operation of a magneto-optical trap and the subsequent stable trapping of single atoms in the tweezer traps place stringent requirements on the vacuum, with pressures that must be well below $1\times\,10^{-7}\,$mbar.
    It is thus crucial to establish that no significant out-gassing of the polymers and adhesives used in the manufacturing process takes place. Moreover, for more complex designs, like the TIR lens discussed above, virtual leaks from the volume between the faces must be ruled out. 
    \item Second, previous applications of the fiber technology did mostly take place in liquids or in air~\cite{Asadollahbaik2020}. In other cases, e.g., when the devices where used for endoscopic applications in medicine, the optics where shielded with a sheath or a cover~\cite{Li2020}. On the other hand, alkali atoms are known to strongly corrode many materials. The durability of the structures under long-term exposure to alkali vapors thus has to be established. 
    \item Third, for technological applications the fiber lenses need to be robust enough to remain attached to the fibers even during vacuum pumpdown, a transport of the experimental setup or other vibrations.
\end{itemize}

Our experimental apparatus to perform these tests consists of a 6-inch stainless steel cube that is permanently pumped by a $50\,$L/s ion pump to reach a steady-state pressure of approximately $10^{-9}\,$mbar without any baking. These vacuum conditions are maintained over several months of operation, with no detectable degradation, outgassing or leaking. Six viewports are used to provide optical access for the laser beams required to form a magneto-optical trap (MOT) of ${}^{85}$Rb. 

\begin{figure}[tb]
\includegraphics[width=0.46\textwidth]{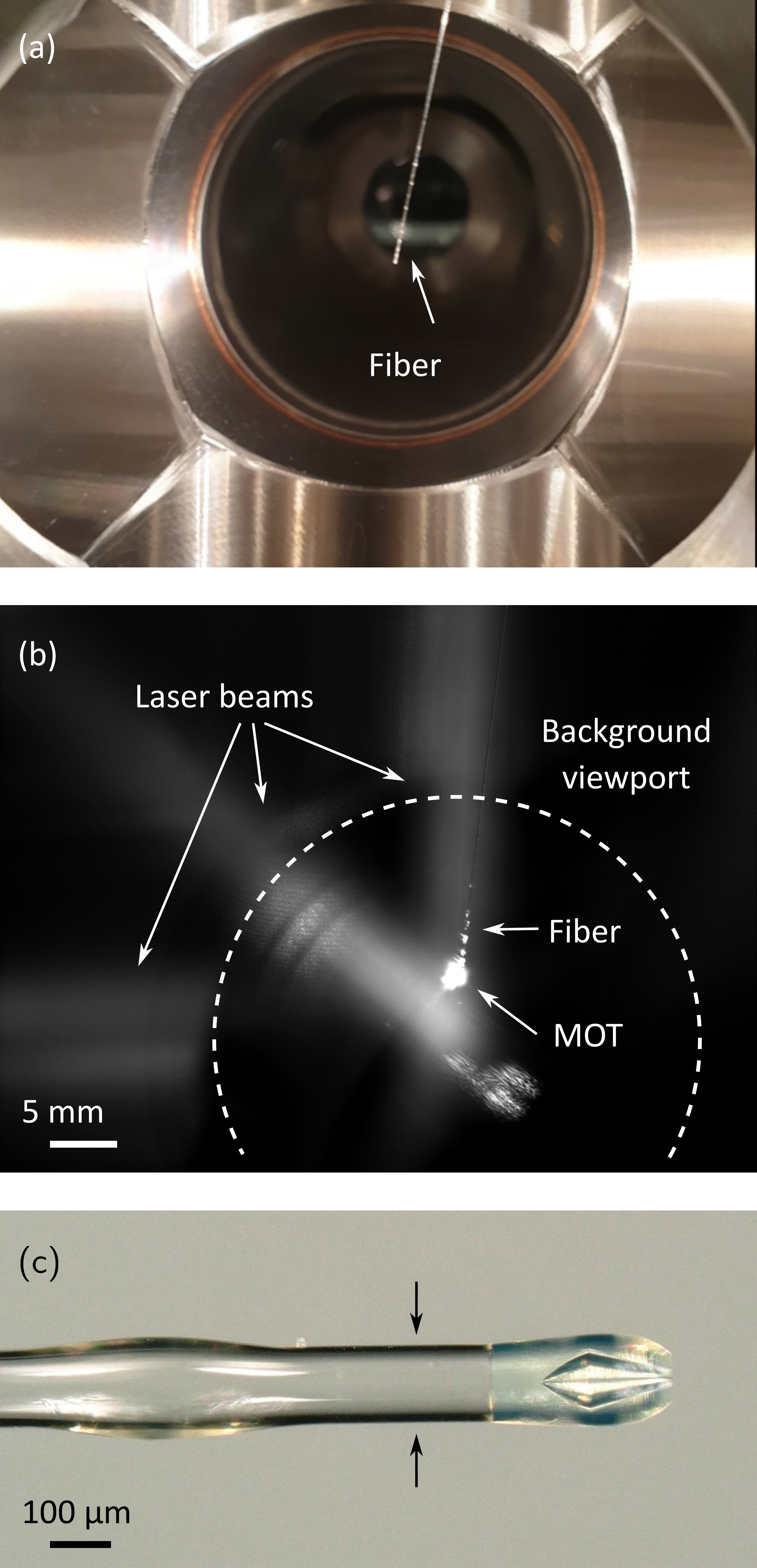}
\caption{(a) Positioning of the fiber trap in the center of a UHV chamber. (b) Creation of a magneto-optical trap $\sim 125\,\mu$m from the fiber tip. (c) TIR lens and fiber tip after continuous long-term exposure to high-pressure rubidium vapor, and subsequent removal from the vacuum chamber. The arrow mark a faint black coating that is due to rubidium oxide. Test images similar to the ones shown in Fig.~\ref{fig:potentials}a,b confirm that this does not affect the performance of the lens.}
\label{fig:experiment}
\end{figure}

The fibers are installed under an angle of $22$ degrees relative to the vertical beam axis of the vacuum chamber, with their fiber tips including the micro-optics positioned in the center of the chamber (see Fig.~\ref{fig:experiment}a). A standard Swagelok fitting in combination with a teflon ferrule is used to feed the fiber into vacuum~\cite{Abraham1998}. The inner diameter of the teflon ferrule is $300\,\mu$m in diameter, such that the $125\,\mu$m diameter fiber can simply pulled through the ferrule before inserting it into the Swagelok fitting and the chamber. We observe no indication that the pressure in the vacuum chamber is affected by the presence of the various fiber trap designs. 

To test the interaction of the fiber with highly-reactive alkali vapors, we intentionally expose the fibers to high rubidium pressures over several hours. We do not observe any influence of this on the optics, except for a thin layer of rubidium oxide forming after removing the fiber from the vacuum chamber Fig.~\ref{fig:experiment}c.
 
The printed structures are further robust against vibrations during mounting and operation of the experiment. To test this, the experimental setup was moved several times on a conventional, undamped table, through elevators, up steps and between rooms without air conditioning. Despite significant vibrations with short-term linear accelerations of up to $20\,$m/$\mathrm{s}^2$ and significant temperature changes, the lenses remain perfectly attached to the fibers and no negative effect on their optical performance could be observed.

Finally, we realize magneto-optical trapping of ultracold atoms near the fiber tip. The trap is a macroscopic six-beam MOT, operated with a gradient of $10\,$G/cm and combined cooling and repumping laser powers of approximately $1.4\,$mW per $6\,$mm diameter beam. Its position relative to the tip of the fiber can be manipulated using additional offset coils. We observe this position by imaging the atomic fluorescence using two CCD cameras positioned perpendicular to each other, and rotated by $45$ degrees with respect to the MOT beam axes. With this, we determine that we successfully trap approximately $10^6$ atoms, at temperatures close to the Doppler limit, and with the center of the MOT located around $125\,\mu$m from the room temperature fiber tip (see Fig.~\ref{fig:experiment}b).

\section{Conclusion and Outlook}
We have introduced a new optical tweezer concept based on tailored, 3D printed lenses on the tip of optical fibers. The lenses are of microscopic dimensions, flexible in their design, UHV compatible and robust. 

As a next step, we will demonstrate the entire procedure from the dipole trapping of single atoms to their manipulation in the tweezer trap. Currently, background light scatter from the large-diameter laser beams used for the demonstration of the macroscopic magneto-optical trap renders the detection of single trapped atoms in the fiber trap challenging. This limitation can be fully overcome using integrated and miniaturized setups for ultracold atoms~\cite{McGilligan2020,Burrow2021}. 

Our approach integrates particularly well with microfabricated chip, grating and pyramidal MOT assemblies~\cite{Wildermuth2004,Nshii2013,Wu2017,Bowden2019} and atom chips~\cite{Gallego2009,Heine2010}. We also expect it to be useful for optical manipulation in light-sensitive areas, such as for atoms inside cryostats~\cite{Nirrengarten2006,Bernon2013,Cantat2020,Schymik2021}, in space applications~\cite{Becker2018,Aveline2020}, as well as inside high-field regions with restricted optical access, as required e.g. for certain types of precision measurements~\cite{Altuntas2018,Kogel2021}. 
It will further be interesting to combine the approach with additive manufacturing techniques for cold atom setups~\cite{Saint2018,Madkhaly2021}. 

A benchmark application of such a microscopic setup will be the realization of a source of indistinguishable single photons. The polymers used in this work are also compatible with $1310\,$nm and $1550\,$nm light, which opens the possibility of operating such a source at telecom wavelengths. In rubidium atoms, this could be realized by using a transition between the $5\,\mathrm{P}_{1/2}$ and $4\,\mathrm{D}_{3/2}$ states~\cite{Uphoff2015}. Furthermore, using multicore fibers it is also possible to integrate arrays of traps into one fiber. In combination with a suitable atomic species, such as strontium, this could be used to realize single-atom clock arrays~\cite{Madjarov2019,Young2020}.

Finally, even completely passive vacuum setups are conceivable~\cite{Rushton2014}, making these devices fully portable and scalable, and, thus, a promising tool for atomic quantum technologies.
\newpage
\section*{Acknowledgments}
We are indebted to Tilman Pfau for generous support and thank Moritz Berngruber, Max M\"ausezahl, Kevin Ng, Jan-Niklas Schmidt and Artur Skljarow for technical assistance. This project has received funding from the European Research Council (ERC) under the European Union’s Horizon 2020 research and innovation programme (Grant agreements No. 949431 and No. 862549), Vector Stiftung, the RiSC programme of the Ministry of Science, Research and Arts Baden-W\"urttemberg, Carl Zeiss Foundation, Deutsche Forschungsgemeinschaft
(GRK2642), and the Terra Incognita programme of the University of Stuttgart. 

\bibliography{bib}

\end{document}